\documentclass[prd, amsfonts, twocolumn, nofootinbib, showpacs]{revtex4-2}
\usepackage{graphicx, epsfig}
\usepackage{color}
\usepackage{amsmath}
\usepackage{amssymb}
\usepackage{physics}
\newcommand{\be}{\begin{equation}}
\newcommand{\ee}{\end{equation}}
\newcommand{\bea}{\begin{eqnarray}}
\newcommand{\eea}{\end{eqnarray}}

\newcommand{\gapp}{\mathrel{\raise.3ex\hbox{$>$}\mkern-14mu
\lower0.6ex\hbox{$\sim$}}}
\newcommand{\lapp}{\mathrel{\raise.3ex\hbox{$<$}\mkern-14mu
\lower0.6ex\hbox{$\sim$}}}
\def\bbox{{\,\lower0.9pt\vbox{\hrule \hbox{\vrule height 0.2 cm
\hskip 0.2 cm \vrule  height 0.2 cm}\hrule}\,}}

\begin{document}
\title{The spin correlation of fermion pairs created by a Kerr black hole gravitational potential }
\author{De-Chang Dai$^{1,2}$}
\affiliation{ $^1$ Department of Physics, national Dong Hwa University, Hualien, Taiwan, Republic of China}
\affiliation{ $^2$ CERCA, Department of Physics, Case Western Reserve University, Cleveland OH 44106-7079}


\begin{abstract}
\widetext
We study the properties of massive fermions created and scattered by a rotating Kerr black hole. The helicities of the scattered fermions can vary during propagation. A fermion with a right-handed helicity can become either right or left-handed after interacting with the gravitational potential. This implies that measuring characteristics of an escaping particle is insufficient to reconstruct all the characteristics of its infalling partner.  This further means the helicities of a particle pair created by the gravitational potential are not fully entangled. Since spin and helicity share many common features, it is likely that the same is true for spins of spontaneously created particles. 
\end{abstract}


\maketitle

\section{introduction}

\begin{figure}[h]
\includegraphics[width=3cm]{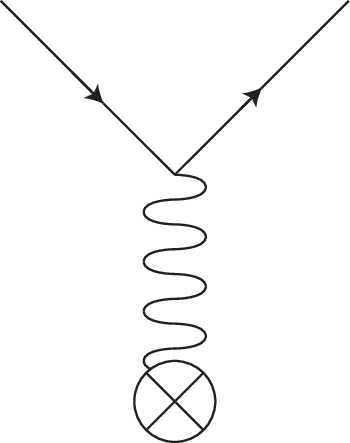}
\caption{ A particle pair is created by an external field. The crossed circle represents the external field. The two upper lines represent two particles created by the external field. This external field cannot be neglected, and it can potentially change these particles' quantum numbers.  
}
\label{pair}
\end{figure}

After Sauter, Heisenberg, and Euler established the action of an electron in a constant electric field\cite{1931ZPhy...69..742S,Heisenberg:1936nmg}, Schwinger realized that charged particles can be created if the electric field is stronger than the critical electric field, $E_{cri}=\frac{m_e c^2}{q_e \hbar}$\cite{Schwinger:1951nm,Feynman:1949}. These particle pairs (Schwinger pairs) originate as virtual particles, and are separated by an external field in order to become a real particle pair. These particle pairs are assumed to be highly entangled since virtual particle pairs must conserve quantum numbers. Spin is generally used to describe this highly entangled state, and the particle's state is written as one of the maximally entangled Bell states  
\begin{equation}
\frac{1}{\sqrt{2}}\ket{\uparrow\downarrow}\pm \frac{1}{\sqrt{2}}\ket{\downarrow\uparrow}.
\end{equation}
This expectation, however, is proved to be incorrect since the external field is involved in the particle creation and can alter the particles' spins\cite{Dai:2019nzv}. The actual creation is illustrated in fig.\ref{pair}. A particle pair is created by an external field, which can alter the characteristics of the particle states. Even though the quantum numbers are preserved at the time of creation, they can vary after particles propagate away from the creation location under the influence of the external field. Thus, taking into account the interaction field is crucial.

Apart from particles created by a strong electromagnetic field, a strong gravitational field can also be a source of particle creation. The best known phenomenon is Hawking radiation\cite{Hawking:1974rv,Hawking:1975vcx}. Hawking radiation is generally considered to be the consequence of particle pairs created by gravity at or near the horizon. One member of the pair escapes to the infinity, while the other falls into the horizon. These two particles are also assumed to be highly entangled and carry opposite quantum numbers. Don Page pointed out that the entanglement entropy of the escaping particles will initially increase but will have to decrease near the halfway of the evaporation process, due to the unitarity of an evaporation process\cite{Page:1993wv}. This is now widely referred to as the Page curve. If these particles are highly entangled, the escaping particles must carry the information of the black hole and can be related to the black hole information paradox problem\cite{Mathur:2009hf}. There are many models and arguments based on the assumption that vacuum particle pairs are maximally entangled\cite{Almheiri:2020cfm}. However, this argument is based on the fact that particle pairs will not interact with the environment after creation. This assumption may not be held since the particle pairs must be ultimately separated by a significant distance. And we know that the entangled pair can be disentangled after propagating some distance\cite{Martin-Martinez:2010yva} or in some other circumstances\cite{Syu:2020btq}. 

It has been noticed that helicity can change in a curved space\cite{Bruggen:1998pp,Mashhoon,1991PhRvL..66.1259C, Anandan,Friedrich}. This article demonstrates that the helicities of massive fermion pairs generated by a Kerr black hole are not entirely entangled. Here, the helicity is defined by a static observed in the Kerr spacetime. A right-helicity fermion can have either a right- or left-helicity partner. This implies that the assertion that escaping particles carry the same amount of information as the particles falling into the black hole is incorrect. In the following, we first review the fermion in a Kerr space and then use a perturbation theory to study how a massive fermion changes its helicity while propagating through space.

\begin{figure}[h]
\includegraphics[width=8cm]{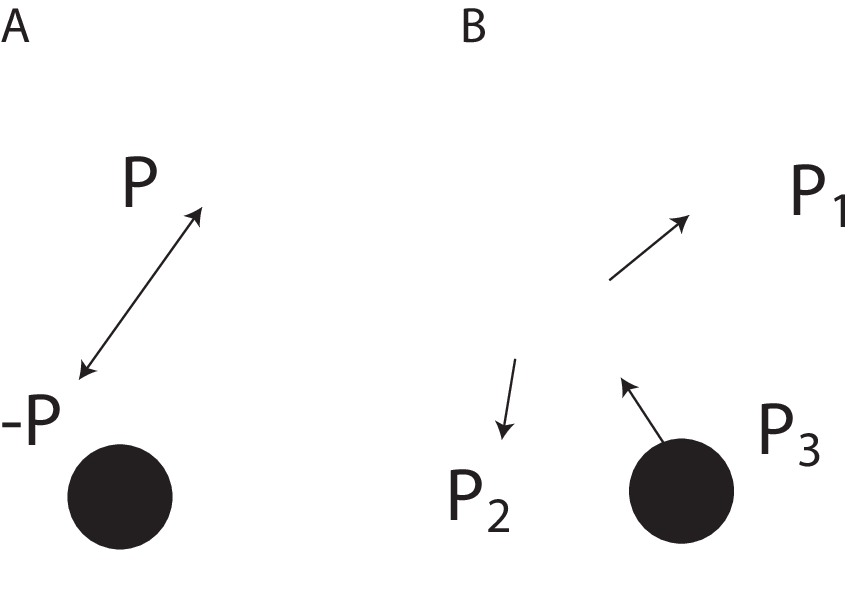}
\caption{Two particles are separated outside a black hole. A: These two particles have opposite momenta at the beginning. B: However, the momenta have not clear correlations after propagating a while. The momentum conservation must be applied to the whole system ($\mathbf{p_1}$, $\mathbf{p_2}$, and $\mathbf{p_3}$).        
}
\label{momentum}
\end{figure}

We notice that many quantities may be highly correlated initially, but the correlations will not be preserved as particles propagate away from the location they are created. The simplest one is momentum. When a particle pair is created, one of them has a momentum $\mathbf{p}$ and the other $-\mathbf{p}$ (Fig. \ref{momentum}). After they propagate away from the creation location, their momenta become $\mathbf{p_1}$ and $\mathbf{p_2}$. The correlation is gone since $\mathbf{p_3}$ is involved. The momentum conservation initially applied to these two particles, must be applied now to the full system, i.e. the black hole and these two particles.   

\section{Fermions in the Kerr black hole background}

Propagation of  massive fermions in a rotation black hole background is a well studied problem. Here, we review only the required background. Detailed calculations can be found in \cite{Chandrasekhar_book}. 

The geometry of a rotating black hole is described by the Kerr metric,
\begin{eqnarray}
ds^2 &=& -(1-\frac{2Mr}{\Sigma})dt^2 -\frac{4Mra\sin^2\theta}{\Sigma} dt d\phi +\frac{\Sigma}{\Delta}dr^2 \nonumber\\
&+&\Sigma d\theta^2 +\Big(r^2+a^2 +\frac{2Mra^2\sin^2\theta}{\Sigma}\sin^2\theta \Big) d\phi^2,\\
\Delta &=& r^2 -2Mr +a^2,\\
\Sigma&=&r^2 +a^2\cos^2\theta ,
\end{eqnarray}
where $a$ is the black hole rotation parameter. The radius of the horizon, $r_h$, is the root of $\Delta=0$.

The massive Dirac equation without an (non-gravitational) external potential is

\begin{equation}
(\gamma^\mu \nabla _\mu +im) \Psi =0,
\end{equation}
where $\gamma^\mu$ are the general relativistic Dirac matrices, which satisfy
\begin{equation}
\{\gamma^\mu,\gamma^\nu \}=2g^{\mu\nu} . 
\end{equation}
 The metric connection is 
\begin{equation} 
\nabla_\mu =\partial_\mu +\frac{1}{8}\omega_{\mu\alpha\beta} [\gamma^\alpha,\gamma^\beta ] 
\end{equation}
where $\omega_{\mu\alpha\beta}$ is the spin connection. The Dirac field, $\Psi$, is a 4-spinor. It can be written in the chiral-2 spinor representation, 
\begin{equation}
\Psi = \begin{bmatrix}
       P^A       \\[0.3em]
       \bar{Q}_{B'}   
     \end{bmatrix},
\end{equation}
and $\gamma^\mu$ is 
\begin{equation}
\gamma^\mu =\sqrt{2}\begin{bmatrix}
       0_{C^2}  & \sigma^{\mu AB'}     \\[0.3em]
       \sigma^\mu _{AB'} & 0_{C^2}   
     \end{bmatrix}
\end{equation}
where $P^A$ and $\bar{Q}_{B'}$ denote 2-component spinors, while $\sigma^\mu_{AB'}$ are the Hermitian ($2\times 2$)-Infeld-van der Waerden symbols. $A\in {1,2}$ and  $B'\in {1',2'}$. The 2-spinor form of the Dirac equation is 
\begin{eqnarray}
\sigma^\mu_{AB'}\nabla_\mu P^A +\frac{im}{\sqrt{2}} \bar{Q}_{B'} =0&\\
\sigma^\mu_{AB'}\nabla_\mu Q^A +\frac{im}{\sqrt{2}} \bar{P}_{B'} =0. 
\end{eqnarray}
here, 
 
\begin{equation}
\sigma^\mu _{(k)(l')} = \begin{bmatrix}
       l^\mu  & m^\mu     \\[0.3em]
       \bar{m}_\mu & n^\mu   
     \end{bmatrix}
\end{equation}
and the null vectors are chosen to be 
\begin{eqnarray}
l^\mu&=&\frac{1}{\Delta}(r^2+a^2,\Delta,0,a)\\
n^\mu&=&\frac{1}{2\rho^2}(r^2+a^2,-\Delta,0,a)\\
m^\mu&=&\frac{1}{\bar{\rho} \sqrt{2}}(ia\sin\theta,0,1,i\csc\theta)\\
\bar{m}^\mu &=& m^{\mu*}
\end{eqnarray}
Here, $\bar{\rho}=r+ia\cos\theta$. These 2-spinors Dirac spinors can be written as 
\begin{eqnarray}
\label{mv-1}
P^0&=&\frac{e^{-i\omega t +ik\phi }}{\sqrt{2}(r-ia\cos\theta)}f_1\\
\label{mv-2}
P^1&=&e^{-i\omega t +ik\phi } f_2\\
\label{mv-3}
\bar{Q}_{0'}&=&e^{-i\omega t +ik\phi } g_1\\
\label{mv-4}
\bar{Q}_{1'}&=&\frac{e^{-i\omega t +ik\phi }}{\sqrt{2}(r+ia\cos\theta)}g_2
\end{eqnarray}
$k\in Z+\frac{1}{2} $.  The Dirac equation reduces to four equations
\begin{eqnarray}
\label{feq-1}
D_0 f_1 +L_{\frac{1}{2}}f_2&=&(i m r +a m \cos\theta)g_1\\
\label{feq-2}
\Delta D_{\frac{1}{2}}^{\dagger} f_2 -L_{\frac{1}{2}}^\dagger f_1&=&-(im r +am \cos\theta)g_2\\
\label{feq-3}
D_0 g_2 -L_{\frac{1}{2}}^\dagger g_1 &=&(im r -am \cos\theta)f_2\\
\label{feq-4}
\Delta D_{\frac{1}{2}}^{\dagger} g_1 +L_{\frac{1}{2}} g_2&=&-(i m r -am \cos\theta)f_1
\end{eqnarray}
Here, 
\begin{eqnarray}
D_n&=&=\partial_r +\frac{iK}{\Delta}+2n\frac{r-M}{\Delta}\\
D_n^\dagger&=&=\partial_r -\frac{iK}{\Delta}+2n\frac{r-M}{\Delta}\\
L_{\frac{1}{2}} &=&\partial_\theta +Q+\frac{1}{2}\cot\theta\\
L_{\frac{1}{2}}^\dagger &=&\partial_\theta -Q+\frac{1}{2}\cot\theta\\
K&=&-(r^2+a^2)\omega +ak\\
Q&=&-a\omega \sin\theta +k\csc\theta
\end{eqnarray}

One may notice that 
\begin{eqnarray}
\Delta^{\frac{1}{2}} D_{\frac{1}{2}}&=&D_0 \Delta^{\frac{1}{2}}\\
\Delta^{\frac{1}{2}} D_{\frac{1}{2}}^{\dagger}&=&D_0^{\dagger} \Delta^{\frac{1}{2}}
\end{eqnarray}

These relationships are applied to simplify some equations. 



\section{Massless fermion case}
It is known that a massless fermion does not change its helicity. The helicities of particle pairs created by spontaneous vacuum decay are always entangled with their companion particles. Therefore, we have to ultimately consider massive particles. A massless fermion's chirality eigenstate is also its helicity eigenstate. This characteristic makes it a good zeroth-order approximation to the massive fermion helicity eigenstates. Here, the solution for massless fermion is reviewed. A detailed review can be found in \cite{Chandrasekhar_book} and related calculation\cite{Dai:2023ewf,Dai:2023zcj}. 

Massless fermions have two chirality states, a right-chirality and left-chirality. The chirality states happen to be their helicity states. We note that the right-chirality states coincide with the right-helicity states, and they coincide with the anti-particles' left-helicity states, and vice versa. One may notice that helicity depends on the relative velocity between particles and observers. However, the helicities of the particles are usually studied in the static frame.We are concerned only with particles at infinity and the horizon, where helicity coincides with the spin in the radial direction.  The radial spin component will not change with boosts in the radial direction and is conserved while these particles are created. Therefore, without the loss of generality helicity is considered only in the static frame in this article instead of an arbitrary frame.  

The right-handed chiral states are written as 
\begin{eqnarray}
\label{mass-1}
P^0&=&\frac{e^{-i\omega t +ik\phi }}{\sqrt{2}(r-ia\cos\theta)}H^{-}_{k\lambda}(r)S^{-}_{k\lambda}(\theta)\\
\label{mass-2}
P^1&=&e^{-i\omega t +ik\phi } H^{+}_{k\lambda}(r)S^{+}_{k\lambda}(\theta)\\
\label{mass-3}
\bar{Q}_{0'}&=&0\\
\label{mass-4}
\bar{Q}_{1'}&=&0
\end{eqnarray}
This represents a right-helicity fermion or a left-helicity anti-fermion.  The left-handed chiral states are written as 
\begin{eqnarray}
\label{mass-5}
P^0&=&0\\
\label{mass-6}
P^1&=&0\\
\label{mass-7}
\bar{Q}_{0'}&=&e^{-i\omega t +ik\phi } H^{+}_{k\lambda}(r)S^{-}_{k\lambda}(\theta)\\
\label{mass-8}
\bar{Q}_{1'}&=&\frac{e^{-i\omega t +ik\phi }}{\sqrt{2}(r+ia\cos\theta)}H^{-}_{k\lambda}(r)S^{+}_{k\lambda}(\theta)
\end{eqnarray}
This represents a left-helicity fermion or a right-helicity anti-fermion. $H^{\pm }_{k\lambda}$ and $S^{\pm }_{k\lambda}$ satisfy

\begin{eqnarray}
\label{couple-1}
\Delta^{\frac{1}{2}}D_0 H^{-}_{k\lambda}&= & \lambda \Delta^{\frac{1}{2}}H^{+}_{k\lambda}\\
\label{couple-2}
\Delta^{\frac{1}{2}}D_0^\dagger \Delta ^{\frac{1}{2}}H^{+}_{k\lambda}&= & \lambda H^{-}_{k\lambda}\\
\label{couple-3}
L_{\frac{1}{2}}S^{+}_{k\lambda}&=&-\lambda  S^{-}_{k\lambda}\\
\label{couple-4}
L_{\frac{1}{2}}^\dagger S^{-}_{k\lambda}&=&\lambda  S^{+}_{k\lambda}
\end{eqnarray}
$\lambda$ is a constant related to the total angular momentum. 

\begin{figure}[h]
\includegraphics[width=8cm]{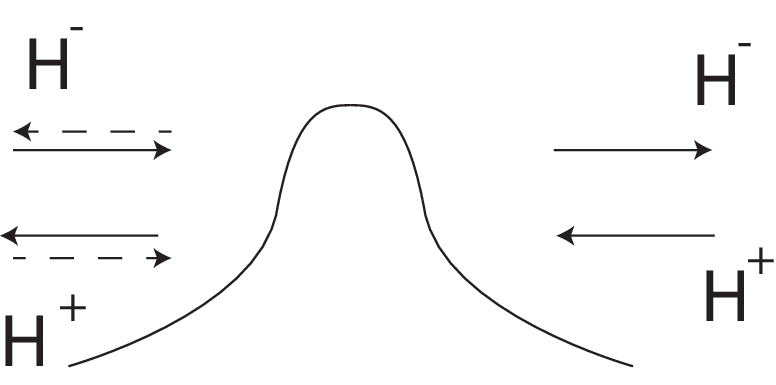}
\caption{$H^+_{k\lambda}$ represents a particle propagating from infinity to the black hole (incoming mode) and a particle propagating toward the horizon (down mode).  $H^-_{k\lambda}$ represents a particle propagating from the black hole to the infinity(outgoing mode) and a particle escaping from the horizon (up mode).  The dash arrows represent that if the propagation nearby the horizon changes its direction if $\omega<k\Omega$.  
}
\label{massless}
\end{figure}

As $r\rightarrow \infty$, 
\begin{eqnarray}
\Delta^{\frac{1}{2}}H^{+}_{k\lambda}&\sim &e^{-i\omega r}\\
H^{-}_{k\lambda}&\sim &e^{i\omega r}.
\end{eqnarray}
, $H^{+}_{k\lambda}$ is the incoming mode, which propagates from infinity toward the black hole, and $H^{-}_{k\lambda}$ is the outgoing mode, which propagates from the black hole toward infinity (Fig. \ref{massless}). 

As $r\rightarrow r_h$, 
\begin{eqnarray}
\Delta^{\frac{1}{2}}H^{+}_{k\lambda}&\sim & e^{-i\omega^* r^*}\\
H^{-}_{k\lambda}&\sim &e^{i\omega^* r^*}, 
\end{eqnarray}
$H^{+}_{k\lambda}$ is the down mode, which propagates toward the horizon, and $H^{-}_{k\lambda}$ is the up mode, which propagates away from the horizon (Fig. \ref{massless}). $r^*$ is a redefined radius parameter,    
\begin{equation}
dr^*=\frac{r^2+a^2}{\Delta} dr ,
\end{equation}

We focus on two types of solutions in this study. The first one is a wave coming from $r\rightarrow \infty$. A part of this wave is reflected back by the gravitational potential, while the rest of it crosses the gravitational barrier. The asymptotic behavior is 

\begin{eqnarray}
\label{in-1}
\Delta^{\frac{1}{2}}H^{+}_{k\lambda}&\rightarrow & \frac{1}{\sqrt{\omega}}e^{-i\omega r} \text{, as } r\rightarrow \infty\\
\label{in-2}
\Delta^{\frac{1}{2}}H^{+}_{k\lambda}&\rightarrow &  \frac{T_{k\lambda}}{\sqrt{\omega^*}}e^{-i\omega^* r^*} \text{, as } r\rightarrow r_h \\
\label{in-3}
H^{-}_{k\lambda}& \rightarrow &  \frac{R_{k\lambda}}{\sqrt{\omega}}e^{i\omega r} \text{, as } r\rightarrow \infty\\
\label{in-4}
H^{-}_{k\lambda}&\rightarrow &0 \text{, as } r\rightarrow r_h
\end{eqnarray} 
$T_{k\lambda}$ and $R_{k\lambda}$ are the transmission factor reflected factor respectively. 
The second one is a wave originating from $r\rightarrow r_h$. Part of the wave is reflected by the gravitational potential, and the rest of it crosses the gravitational barrier. The asymptotic behavior is 
\begin{eqnarray}
\label{out-1}
\Delta^{\frac{1}{2}}H^{+}_{k\lambda}&\rightarrow & 0 \text{, as } r\rightarrow \infty\\
\label{out-2}
\Delta^{\frac{1}{2}}H^{+}_{k\lambda}&\rightarrow &  \frac{r_{k\lambda}}{\sqrt{\omega^*}}e^{-i\omega^* r^*} \text{, as } r\rightarrow r_h \\
\label{out-3}
H^{-}_{k\lambda}& \rightarrow &  \frac{t_{k\lambda}}{\sqrt{\omega}}e^{i\omega r} \text{, as } r\rightarrow \infty\\
\label{out-4}
H^{-}_{k\lambda}&\rightarrow &  \frac{1}{\sqrt{\omega^*}}e^{i\omega^* r^*}  \text{, as } r\rightarrow r_h
\end{eqnarray} 
$t_{k\lambda}$ and $r_{k\lambda}$ are the transmission factor reflected factor respectively. 

These two solutions are the zeroth-order approximation solutions for the massive fermions.

\section{Massive fermions}
We adopt now the perturbation theory and choose the mass-related terms in eq (\ref{feq-1}) to (\ref{feq-4}) as the perturbation terms. We considered first the right-handed massless fermion as a zeroth-order approximation, and the solution is 
\begin{eqnarray}
f_1&=&H^{-}_{k\lambda}(r)S^{-}_{k\lambda}(\theta) +O(m^2)\\
f_2&=&H^{+}_{k\lambda}(r)S^{+}_{k\lambda}(\theta) + O(m^2)\\
g_1&=&g_2=O(m)
\end{eqnarray}
$O(m^n)$ means that the lowest order correction terms are proportional to $m^n$. $H^{-}_{k\lambda}(r)$ and $H^{+}_{k\lambda}(r)$ are the same as eq. (\ref{in-1}) to eq. (\ref{in-4}).

\begin{eqnarray}
D_0 g_2 -L_{\frac{1}{2}}^\dagger g_1 &=&(im r -am \cos\theta)H^{+}_{k\lambda}S^{+}_{k\lambda}\\
\Delta D_{\frac{1}{2}}^{\dagger} g_1 +L_{\frac{1}{2}} g_2&=&-(i m r -am \cos\theta)H^{-}_{k\lambda}S^{-}_{k\lambda}
\end{eqnarray}


Since $S^{-}_{k\lambda}$ and $S^{+}_{k\lambda}$ are both a complete set, the following terms can be decomposed to  
\begin{eqnarray}
\cos\theta S^{+}_{k\lambda}&=&\sum_{\lambda_1} a^{-}_{k\lambda\lambda_1} S^{-}_{k\lambda_1}\\
\cos\theta S^{-}_{k\lambda}&=&\sum_{\lambda_1} b^{+}_{k\lambda\lambda_1}S^{+}_{k\lambda_1}.
\end{eqnarray}
As mentioned, $\lambda_1$ is a total-angular-momentum-related constant. The decomposition implies the particle's angular momentum changes.  
Since 
\begin{equation}
S^{-}_{k\lambda}(\theta)=S^{+}_{k\lambda}(\pi-\theta)
\end{equation}
, then 
\begin{equation}
a^{-}_{k\lambda\lambda_1}=-b^{+}_{k\lambda\lambda_1}.
\end{equation}
The decompositions are chosen to solve the equations easily. $g_1$ and $g_2$ can be decomposed to 
\begin{eqnarray}
g_1&=&\sum_{\lambda_1} e^{+}_{k\lambda\lambda_1} S^-_{k\lambda_1}\\
g_2&=& \sum_{\lambda_1} h^{-}_{k\lambda\lambda_1} S^+_{k\lambda_1}
\end{eqnarray}
The $\pm$ signs are chosen to match the symbols in the massless case. 
For $\lambda_1=\lambda$, 
\begin{eqnarray}
\Delta^{\frac{1}{2}} D_0  h^{-}_{k\lambda\lambda} -\lambda \Delta^{\frac{1}{2}}  e^{+}_{k\lambda\lambda}&=&im r\Delta^{\frac{1}{2}}H^{+}_{k\lambda}\\
\Delta^{\frac{1}{2}} D_0^{\dagger}\Delta^{\frac{1}{2}}e^{+}_{k\lambda\lambda} - \lambda  h^{-}_{k\lambda\lambda} &=&-i m r H^{-}_{k\lambda}
\end{eqnarray}
As $r\rightarrow \infty$, the corresponding terms are 
\begin{eqnarray}
h^{-}_{k\lambda\lambda}&\rightarrow& \frac{m }{2\omega}\Delta^{\frac{1}{2}}H^{+}_{k\lambda}\\
\Delta^{\frac{1}{2}}e^{+}_{k\lambda\lambda}&\rightarrow&\frac{ m}{2\omega} H^{-}_{k\lambda}
\end{eqnarray}
As $r\rightarrow r_h$, the 
\begin{eqnarray}
h^{-}_{k\lambda\lambda}&\rightarrow& 0\\
\Delta^{\frac{1}{2}}e^{+}_{k\lambda\lambda}&\rightarrow&0
\end{eqnarray}
This condition holds because, near the horizon, the particle momentum is so high that mass can be neglected.  
This is the first order correction term for a free massive fermion, not related to the scattering process. Therefore, we will not discuss the details of it.  
The more important terms are $\lambda_1\neq \lambda$ terms, which are induced by the gravitational scattering. The corresponding terms are 

\begin{eqnarray}
\Delta^{\frac{1}{2}} D_0  h^-_{k\lambda\lambda_1} -\lambda_1 \Delta^{\frac{1}{2}}  e^+_{k\lambda\lambda_!} &=&-am  a^{-}_{k\lambda\lambda_1}\Delta^{\frac{1}{2}}H^{+}_{k\lambda}\\
\Delta^{\frac{1}{2}} D_0^{\dagger}\Delta^{\frac{1}{2}}e^+_{k\lambda\lambda_1} - \lambda_1  h^-_{k\lambda\lambda_1} &=&+am b^{+}_{k\lambda\lambda_1}H^{-}_{k\lambda}
\end{eqnarray}
By applying equation (\ref{couple-1}) and (\ref{couple-2}), the solution is found,  
\begin{eqnarray}
\Delta^{\frac{1}{2}}e^{+}_{k\lambda\lambda_1} &=&E_{k\lambda\lambda_1} H^{+}_{k\lambda_1}+\frac{am a^{-}_{k\lambda_1\lambda}}{\lambda_1-\lambda} H^{+}_{k\lambda}\\
h^{-}_{k\lambda\lambda_1} &=&E_{k\lambda\lambda_1} H^{-}_{k\lambda_1}-\frac{am b^{+}_{k\lambda_1\lambda}}{\lambda_1-\lambda} H^{-}_{k \lambda}
\end{eqnarray}
$E_{k\lambda_1\lambda}$ can be adjusted  according to the condition.  Note that these are the first order correction terms which are proportional to $m$. More involving calculations are needed to include higher order terms. 

Since $g_1$ and $g_2$ are the left-charility components,  $e^{+}_{k\lambda\lambda_1}$  and $h^{-}_{k\lambda\lambda_1}$ represent the opposite chirality or helicity solutions to the original particles. One can also start from a left-helicity massless fermion as the zeroth-order approximation and find the corresponding solution.  

\section{Asymptotic behavior}
Even though $(P^1,P^2,\bar{Q}_1,\bar{Q}_2)^T$ are the complete components of the wave function, $(f_1,\Delta^{\frac{1}{2}}f_2,\Delta ^{\frac{1}{2}}g_1,g_2)^T$ is a more concise way to represent the required components. Therefore, in the following, we list only $(f_1,\Delta^{\frac{1}{2}}f_2,\Delta ^{\frac{1}{2}}g_1,g_2)^T$. 

\begin{figure}[h]
\includegraphics[width=8cm]{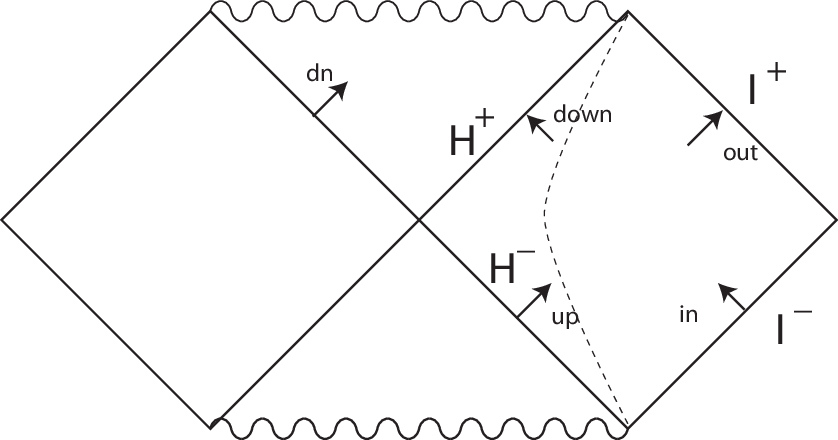}
\caption{The Penrose-Carter diagram: The space outside the horizon is presented in the right square. The upper triangle represents the space inside the future horizon. The black hole radiation due to the Hawking effect involves these two regions. On the other hand, only the right square is involved in particle creation by the superradiance mechanism. The thin dashed line represents the potential barrier that induces the superradiance. Four types of bases ($down$, $up$, $in$, and $out$) are involved in the process. Five bases (including $dn$) are involved in the full black hole radiation \cite{Frolov}. Similar argument for scalar field can be found in \cite{Dai:2023ewf,Dai:2023zcj}.       
}
\label{penrose-carter}
\end{figure}

There are four types of solutions. The right-helicity ones have the following asymptotic behavior,  

\begin{eqnarray}
\psi_{\lambda,R}^{in}&=&\frac{e^{-i\omega r}}{\sqrt{\omega}}S^+_{k\lambda}(0,1,0,\frac{m}{2\omega})^T \\
\psi_{\lambda,R}^{up}&=&\frac{e^{+i\omega^* r^*}}{\sqrt{\omega^*}}S^-_{k\lambda}(1,0,0,0)^T \\
\psi_{\lambda,R}^{out}&=&\frac{e^{+i\omega r}}{\sqrt{\omega^*}}S^-_{k\lambda}(1,0,\frac{m}{2\omega},0)^T  \\
\psi_{\lambda,R}^{down}&=&\frac{e^{-i\omega^* r^*}}{\sqrt{\omega^*}}S^+_{k\lambda}(0,1,0,0)^T 
\end{eqnarray} 
$R$ in $\psi_{\lambda,R}^{\alpha}$  represents the right-helicity, while  index $\alpha$ has the same meaning as Fig. \ref{penrose-carter}.  The left-helicity ones have the following asymptotic behavior,  

\begin{eqnarray}
\psi_{\lambda,L}^{in}&=&\frac{e^{-i\omega r}}{\sqrt{\omega}}S^+_{k\lambda}(0,\frac{m}{2\omega},0,1)^T \\
\psi_{\lambda,L}^{up}&=&\frac{e^{+i\omega^* r^*}}{\sqrt{\omega^*}}S^-_{k\lambda}(0,0,1,0)^T\\
\psi_{\lambda,L}^{out}&=&\frac{e^{+i\omega r}}{\sqrt{\omega^*}}S^-_{k\lambda}(\frac{m}{2\omega},0,1,0)^T\\
\psi_{\lambda,L}^{down}&=&\frac{e^{-i\omega^* r^*}}{\sqrt{\omega^*}}S^+_{k\lambda}(0,0,0,1))^T
\end{eqnarray} 
$L$ in $\psi_{\lambda,L}^{\alpha}$  represents the right-helicity. 

A right-helicity particle ($\sim \psi_{k\lambda,R}^{in}$ ) from infinity is scattered by the the gravitational potential then becomes a combination of other $\psi_{\lambda,\beta}^{\alpha}$, at late time,    
\begin{eqnarray}
\psi_{k\lambda,R}^{in}&\rightarrow& T_{k\lambda} \psi^{down}_{k\lambda,R} +R_{k\lambda} \psi^{out}_{k\lambda,R} \nonumber\\
\label{wave-relation-1}
&+&  \sum_{\lambda_1}T_{k\lambda\lambda_1}^{R,L} \psi^{down}_{k\lambda_1,L} +R_{k\lambda\lambda_1}^{R,L} \psi^{out}_{k\lambda_1,L} 
\end{eqnarray}
Here, $T_{k\lambda}$ and $R_{k\lambda}$ are the same as (\ref{in-2}) and (\ref{in-3}) at least to $O(m^2)$. $T_{k\lambda\lambda_1}^{R,L}$ and $R_{k\lambda\lambda_1}^{R,L}$ can be determined by setting 
\begin{equation}
E_{k\lambda\lambda_1} =\lim_{r\rightarrow \infty}-\frac{am a^{-}_{k\lambda_1\lambda}}{\lambda_1-\lambda}  \frac{H^{+}_{k\lambda}}{H^{+}_{k\lambda_1}}
\end{equation}
here, $H^{\pm}_{k\lambda}$ are from eq. (\ref{in-1}) to (\ref{in-4}). $\psi_{k\lambda,R}^{-\infty}$ is the only incoming mode from infinity. $e^{+}_{k\lambda\lambda_1}$ does not vanish as $r\rightarrow r_h$ since the transmission coefficient for $\lambda$ and $\lambda_1$ modes are not the same ($T_{k\lambda}\neq T_{k\lambda_1}$). The transmission amplitude is 

\begin{equation}
T_{k\lambda\lambda_1}^{R,L}=-\frac{am a^{-}_{k\lambda_1\lambda}}{\lambda_1-\lambda}T_{k\lambda_1} +\frac{am a^{-}_{k\lambda_1\lambda}}{\lambda_1-\lambda}T_{k\lambda}
\end{equation}

  $h^{+}_{k\lambda\lambda_1}$ does not vanish as $r\rightarrow \infty$ since the reflection coefficient for $\lambda$ and $\lambda_1$ modes are not the same ($R_{k\lambda}\neq R_{k\lambda_1}$). The reflection amplitude is 
  
\begin{equation}
R_{k\lambda\lambda_1}^{R,L} = -\frac{am a^{-}_{k\lambda_1\lambda}}{\lambda_1-\lambda}R_{k\lambda_1}-\frac{am b^{+}_{k\lambda_1\lambda}}{\lambda_1-\lambda}R_{k\lambda}
\end{equation}
 
At the same time, a particle escaping from the horizon and scattered by the gravitational potential is described by
\begin{eqnarray}
\psi_{k\lambda,R}^{up}&\rightarrow & r_{k\lambda} \psi^{down}_{k\lambda,R} +t_{k\lambda} \psi^{out}_{k\lambda,R} \nonumber\\
\label{wave-relation-2}
&+&  \sum_{\lambda_1}r_{k\lambda\lambda_1}^{R,L} \psi^{down}_{k\lambda_1,L} +t_{k\lambda\lambda_1}^{R,L} \psi^{out}_{k\lambda_1,L} 
\end{eqnarray}
Here, $t_{k\lambda}$ and $r_{k\lambda}$ are the same as (\ref{out-2}) and (\ref{out-3}) at least to $O(m^2)$. $r_{k\lambda\lambda_1}^{R,L}$, and $t_{k\lambda\lambda_1}^{R,L}$ can be determined by setting 
\begin{equation}
E_{k\lambda\lambda_1} =\lim_{r\rightarrow r_h}\frac{am b^{+}_{k\lambda_1\lambda}}{\lambda_1-\lambda}  \frac{H^{-}_{k\lambda}}{H^{-}_{k\lambda_1}}
\end{equation}
Here, $H^{\pm}_{k\lambda}$ are from eq. (\ref{out-1}) to (\ref{out-4}).
$\psi_{k\lambda,R}^{-\infty}$ is the only outgoing mode from infinity. $e^{+}_{k\lambda\lambda_1}$ does not vanish as $r\rightarrow \infty$ since the transmission coefficient for $\lambda$ and $\lambda_1$ modes are not the same ($t_{k\lambda}\neq t_{k\lambda_1}$). 
\begin{equation}
t_{k\lambda\lambda_1}^{R,L}=\frac{am b^{+}_{k\lambda_1\lambda}}{\lambda_1-\lambda}t_{k\lambda_1} +\frac{am a^{-}_{k\lambda_1\lambda}}{\lambda_1-\lambda}t_{k\lambda}
\end{equation}
 $h^{+}_{k\lambda\lambda_1}$ does not vanish as $r\rightarrow t_h$ since the reflection coefficients for $\lambda$ and $\lambda_1$ modes are not the same ($r_{k\lambda}\neq r_{k\lambda_1}$). 
\begin{equation}
r_{k\lambda\lambda_1}^{R,L} =\frac{am b^{+}_{k\lambda_1\lambda}}{\lambda_1-\lambda}r_{k\lambda_1}-\frac{am b^{+}_{k\lambda_1\lambda}}{\lambda_1-\lambda}r_{k\lambda}
\end{equation}
Because of the symmetry of the right- and left-handed helicity modes, the left-handed helicity modes can be written as  

\begin{eqnarray}
\psi_{k\lambda,L}^{in} &\rightarrow& T_{k\lambda} \psi^{down}_{k\lambda,L} +R_{k\lambda} \psi^{out}_{k\lambda,L} \nonumber\\
\label{wave-relation-3}
&+&  \sum_{\lambda_1}T_{k\lambda\lambda_1}^{L,R} \psi^{down}_{k\lambda_1,R} +R_{k\lambda\lambda_1}^{L,R} \psi^{out}_{k\lambda_1,R} \\
\psi_{k\lambda,L}^{up}&\rightarrow& r_{k\lambda} \psi^{down}_{k\lambda,L} +t_{k\lambda} \psi^{out}_{k\lambda,L} \nonumber\\
\label{wave-relation-4}
&+&  \sum_{\lambda_1}r_{k\lambda\lambda_1}^{L,R} \psi^{down}_{k\lambda_1,R} +t_{k\lambda\lambda_1}^{L,R} \psi^{out}_{k\lambda_1,R} 
\end{eqnarray}

A right-helicity fermion can become either a right- or left-helicity fermion after scattering. The particle's helicity can be changed after the transmission or reflection. It implies that gravitational potential can change the helicity of a particle. Therefore, the helicities of spontaneously created particle pairs may not be completely entangled. The current calculations were performed for helicity, i.e. not for the spin eigenstates, but spin and helicity share many similar properties, so we expect the conclusions to stay the same for spin. 

\section{Bogoliubov transformation }
There are four types of possible wavefunctions which are relevant for us. The asymptotic behavior of these functions is, 
\begin{eqnarray}
\Psi^{in}_{k\lambda, I}&=&\psi^{in}_{k\lambda, I} \text{, as }r\rightarrow \infty\\
\Psi^{out}_{k\lambda, I}&=&\psi^{out}_{k\lambda, I} \text{, as }r\rightarrow \infty\\
\Psi^{up}_{k\lambda, I}&= &\psi^{up}_{k\lambda, I} \text{, as }r\rightarrow r_h\\
\Psi^{down}_{k\lambda, I}&=&\psi^{down}_{k\lambda, I} \text{, as }r\rightarrow r_h
\end{eqnarray}

Fig. \ref{penrose-carter} shows that $\Psi^{in}_{k\lambda, I}$ is a wave starting from infinity and approaching the black hole and getting scattered. $\Psi^{out}_{k\lambda, I}$ is a wave escaping to infinity (originating either from the incoming mode or the horizon).  $\Psi^{up}_{k\lambda, I}$ is a wave escaping from the horizon and getting scattered by the black hole potential. $\Psi^{down}_{k\lambda, I}$ falls into the horizon (originating either from infinity or reflected by the gravitational potential).

Since we need two different wave functions for the Bogoliubov transformation, there are two or more ways to quantize the field 

\begin{eqnarray}
\label{quantize}
\hat{\Psi} &=& \sum_{\omega k\lambda I} \hat{\alpha}_{\omega k\lambda, I}^{in} \Psi^{in}_{k\lambda, I} +\hat{\alpha}_{\omega k\lambda,I}^{up} \tilde{\Psi}^{up}_{k\lambda,I}\nonumber\\
&&+ \hat{\beta}_{\omega k\lambda,I}^{in\dagger} \Psi^{in}_{k\lambda,I}{}^*+  \hat{\beta}_{\omega k\lambda,I}^{ up\dagger} \tilde{\Psi}^{up}_{k\lambda,I}{}^*\\
&=&\sum_{\omega,k,\lambda,I} \hat{\gamma}_{\omega k\lambda,I}^{out} \Psi^{out}_{k\lambda,I} +  \hat{\gamma}_{\omega k\lambda,I}^{ down} \tilde{\Psi}^{down}_{k\lambda,I}\nonumber\\
&&+ \hat{\delta}_{\omega k\lambda,I}^{out\dagger} \Psi^{out}_{k\lambda,I} {}^*+  \hat{\delta}_{\omega k\lambda,I}^{ down\dagger} \tilde{\Psi}^{down}_{k\lambda,I}{}^*
\end{eqnarray}
where, $I$ represents the helicity,  $R$ or $L$. $\hat{\sigma}_{\omega k\lambda, I}^{\zeta}$ are the annihilation operators and $\hat{\sigma}^{\zeta\dagger}_{\omega k\lambda, I}$ are the creation operators. $\alpha$ and $\gamma$ represent the particles, while $\beta$ and $\delta$ represent the antiparticles. 
Also, 
\begin{eqnarray}
 \tilde{\Psi}^{M}_{J,\lambda,I}&= &\Psi^{M}_{J,\lambda,I}\text{, if }\omega - k\Omega_H>0\\
 &= &\psi^{M}_{J,\lambda,I} {}^*\text{, if }\omega - k\Omega_H<0 ,
\end{eqnarray}
where $M$ can be $up$ or $down$. The two types of vacua corresponding to $\hat{\alpha}_{J,\lambda,I}^{M}$,  $\hat{\beta}_{J,\lambda,I}^{M}$, $\hat{\gamma}_{J,\lambda,I}^{M}$ and $\hat{\delta}_{J,\lambda,I}^{M}$ are 
\begin{eqnarray}
\hat{\alpha}_{\omega k\lambda,I}^{M}\ket{in;0}=\hat{\beta}_{\omega k\lambda,I}^{M} \ket{in;0}&=& 0\\
\hat{\gamma}_{\omega k\lambda,I}^{M}\ket{out;0}=\hat{\delta}_{\omega k\lambda,I}^{M} \ket{out;0} &=& 0 .
\end{eqnarray}
where $M$ can be $up$, $down$, $in$ or $out$. We now apply eq. (\ref{wave-relation-1}),  (\ref{wave-relation-2}), (\ref{wave-relation-3}) and    (\ref{wave-relation-4})  to eq. (\ref{quantize}), and find the relationships between the annihilation operators. For $\omega - k\Omega_H>0 $ 
\begin{eqnarray}
\hat{\gamma}^{out}_{\omega k \lambda, I} &=& R_{k\lambda}\hat{\alpha}^{in}_{\omega k \lambda, I}+t_{k\lambda}\hat{\alpha}^{up}_{\omega k \lambda, I} 
\nonumber\\
&+&\sum_{\lambda_1}R_{k\lambda_1 \lambda}^{JI}\hat{\alpha}^{in}_{\omega k \lambda_1, J}+t_{k\lambda_1 \lambda}^{JI}\hat{\alpha}^{up}_{\omega k \lambda_1, J}\\
\hat{\delta}^{out\dagger}_{\omega k \lambda, I} &=&R_{k\lambda}^*\hat{\beta}^{in\dagger}_{\omega k \lambda, I}+t_{k\lambda}^*\hat{\beta}^{up\dagger}_{\omega k \lambda, I} \nonumber\\
&+&\sum_{\lambda_1}R_{k\lambda_1 \lambda}^{JI*} \hat{\beta}^{in\dagger}_{\omega k \lambda_1, J}+t_{k\lambda_1 \lambda}^{JI*} \hat{\beta}^{up\dagger}_{\omega k \lambda_1, J}\\
\hat{\gamma}^{down}_{\omega k \lambda, I} &=&T_{k\lambda}\hat{\alpha}^{in}_{\omega k \lambda, I}+r_{k\lambda}\hat{\alpha}^{up}_{\omega k \lambda, I} 
\nonumber\\
&+&\sum_{\lambda_1}T_{k\lambda_1 \lambda}^{JI}\hat{\alpha}^{in}_{\omega k \lambda_1, J}+r_{k\lambda_1 \lambda}^{JI}\hat{\alpha}^{up}_{\omega k \lambda_1, J}\\
\hat{\delta}^{down}_{\omega k \lambda, I}{}^\dagger &=&T_{k\lambda}^*\hat{\beta}^{in\dagger}_{\omega k \lambda, I}+r_{k\lambda}^*\hat{\beta}^{up\dagger}_{\omega k \lambda, I} \nonumber\\
&+&\sum_{\lambda_1}T_{k\lambda_1 \lambda}^{JI*}\hat{\beta}^{in\dagger}_{\omega k \lambda_1, J}+r_{k\lambda_1 \lambda}^{JI*} \hat{\beta}^{up\dagger}_{\omega k \lambda_1, J}
\end{eqnarray}

$J$ and $I$ represent the helicities of the particles, and $J\neq I$. There is no mixing between annihilation and creation operators. Therefore there is no particle creation outside the horizon. However, particles created by the tunneling through the black hole horizon are still affected by the gravitational potential and change their helicities during their escape to infinity. Therefore, the escaping particles can have a different helicity from their partners inside the black hole.  

For $\omega - k\Omega_H<0$,

\begin{eqnarray}
\hat{\gamma}^{out}_{\omega k \lambda, I} &=& R_{k\lambda}\hat{\alpha}^{in}_{\omega k \lambda, I}+t_{k\lambda}^*\hat{\beta}^{up\dagger}_{-\omega -k \lambda, I} 
\nonumber\\
&+&\sum_{\lambda_1}R_{k\lambda_1 \lambda}^{JI}\hat{\alpha}^{in}_{\omega k \lambda_1, J}+t_{k\lambda_1 \lambda}^{JI*}\hat{\beta}^{up\dagger}_{-\omega -k \lambda_1, J}\\
\hat{\delta}^{out\dagger}_{\omega k \lambda, I} &=&R_{k\lambda}^*\hat{\beta}^{in\dagger}_{\omega k \lambda, I}+t_{k\lambda}\hat{\alpha}^{up}_{-\omega -k \lambda, I} \nonumber\\
&+&\sum_{\lambda_1}R_{k\lambda_1 \lambda}^{JI*} \hat{\beta}^{in\dagger}_{\omega k \lambda_1, J}+t_{k\lambda_1 \lambda}^{JI} \hat{\alpha}^{up}_{-\omega -k \lambda_1, J}{}\\
\hat{\delta}^{down}_{-\omega -k \lambda, I}{}^\dagger  &=&T_{k\lambda}\hat{\alpha}^{in}_{\omega k \lambda, I}+r_{k\lambda}{}^*\hat{\beta}^{up}_{-\omega -k \lambda, I}{}^\dagger \nonumber\\
&+&\sum_{\lambda_1}T_{k\lambda_1 \lambda}^{JI}\hat{\alpha}^{in}_{\omega k \lambda_1, J}+r_{k\lambda_1 \lambda}^{JI*}\hat{\beta}^{up\dagger}_{-\omega -k \lambda_1, J}\\
\hat{\gamma}^{down}_{-\omega -k \lambda, I}&=&T_{k\lambda}^*\hat{\beta}^{in\dagger}_{\omega k \lambda, I}+r_{k\lambda}\hat{\alpha}^{up}_{-\omega -k \lambda, I} \nonumber\\
&+&\sum_{\lambda_1}T_{k\lambda_1 \lambda}^{JI*} \hat{\beta}^{in\dagger}_{\omega k \lambda_1, J}+r_{k\lambda_1 \lambda}^{JI} \hat{\alpha}^{up}_{-\omega -k \lambda_1, J}{}
\end{eqnarray}
Annihilation and creation operators are mixing, and there is particle creation outside the black hole. We are looking for the helicity correlation of the particle pairs. The amplitude of these channels are 
\begin{eqnarray}
A_{I\lambda J \lambda_1} &=&\bra{out;0} \hat{\delta}^{down}_{-\omega -k \lambda, I}\hat{\gamma}^{out}_{\omega k \lambda_1, J} U(t) \ket{in;0} \nonumber \\
&\approx  & \bra{in;0} \hat{\delta}^{down}_{-\omega -k \lambda, I}\hat{\gamma}^{out}_{\omega k \lambda_1, J} \ket{in;0} \nonumber\\
&=&  \begin{cases}
   r_{-k\lambda}t_{k\lambda}^*        & \text{if } I=J \text{ and } \lambda=\lambda_1\\
    r_{-k\lambda\lambda_1}^{IJ} t_{k\lambda}^*   &  \text{if } I\neq J \text{ and } \lambda\neq\lambda_1\\
     0   &  \text{ otherwise }
  \end{cases}
\end{eqnarray}
This calculation is a first-order approximation. A method to study the higher order correction may be found in \cite{Dai:2019nzv,Ma1990,Qin}. 
Since $A_{R\lambda R \lambda_1}$,  $A_{R\lambda L \lambda_1}$, $A_{L\lambda R \lambda_1}$, $A_{L\lambda L \lambda_1}$ are not 0, a right-helicity particle can have both left-helicity and right-helicity anti-particle as its companion, and vice versa. Hence, the helicities of particle pairs created by the gravitational potential are not full entangled. The correlation of helicities is changed due to the variance of the total angular momentum ($\lambda$ and $\lambda_1$).

\section{Maximal entanglement of the particle pairs is not always true} 

 In the previous section, we considered only the correlation of the helicity of particle pairs. Here, we want to point out that angular distribution ($S^{\pm}_{k\lambda}(\theta)$) also ruins the correlation and reduces the entanglement of the particle pairs. Therefore, discussion based on the maximally entangled states in the black hole information paradox is problematic. 

\begin{figure}[h]
\includegraphics[width=8cm]{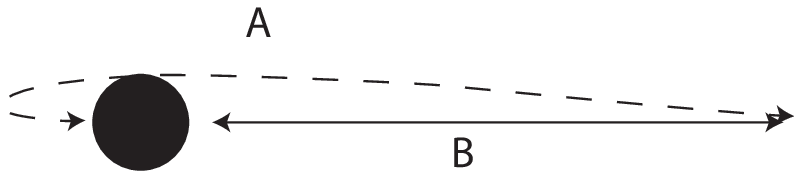}
\caption{ When particle pairs are created, one of the particles in a pair falls into the black hole, and the other escapes to infinity. Some paths of the pairs can be indistinguishable at infinity but are very different near the black hole. For example, a partcle A is absorbed at the opposite side from the partcle B.  
}
\label{path}
\end{figure}

\begin{figure}[h]
\includegraphics[width=8cm]{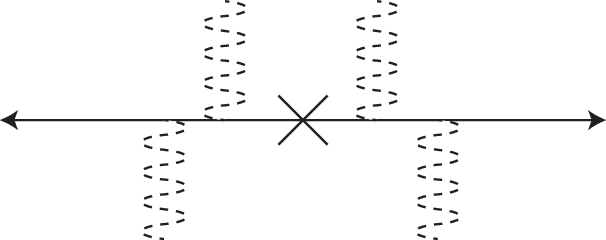}
\caption{  A particle pair is created at X. They interact with the environment while propagating away from X. In this case, they interact through gravity, but they can interact with anything else depending on the conditions. Their physical quantities can change during propagation.        
}
\label{interact}
\end{figure}

A particle is a local concept. The wave function describes the probability of finding particles at a particular position. Fig. \ref{path} shows the paths of two possible particle pairs. The particles at infinity are identical, but the black hole absorbs their corresponding partners at different locations. To simplify the argument, we assume the helicities of the pairs are highly correlated. In this case, the absorbed particles of path A and path B have the same helicity. Their directions of motion are opposite, which implies that their spins are opposite. Even if the helicity is highly correlated, their spin cannot be fully correlated. This means that one cannot predict the corresponding particle's state even if one knows the state of the others. The original assumption that particle pairs share the same amount of information is not always true. Fig. \ref{interact} depicts this process. First, a pair is created. The pair of particles interact with the environment while propagating. Considering that this particle pair is highly correlated is equivalent to neglecting all the interactions in the process. Therefore, it cannot be completely reliable. Most likely, the correlated quantities of the pair should be  
\begin{equation}
\ket{RL} \rightarrow A_1 \ket{RR} + A_2 \ket{RL} +A_3 \ket{LR}+A_4 \ket{LL}
\end{equation}
In this equation, the left side represents the state while the pair is created; the right side represents the state after propagation for a while. The correlation quantity is smoothed out. The density matrix of one of the particles is not diagonalized, and the entanglement entropy can be 
\begin{equation}
\ln(2)\ge S_e\ge 0
\end{equation} 
This result differs from the usual  assumption that the particles must be maximally entangled. It reduces to Hawking's case if the environment eradicates the entanglement. On the other hand, the argument based on maximum entanglement should be redone much more carefully \cite{Page:1993wv,Mathur:2009hf}.  The environment can interrupt the entanglement and hide the information. Therefore, a much more complete discussion is necessary when dealing with the black hole information paradox.

\section{Conclusion}

It is often stated in the literature that vacuum particle pairs created in the black hole background  are fully entangled and should be in one of the Bell states. This assumption, however, has yet to be verified or disputed in the presence of gravity, though many important conclusions are derived from this unverified assumption (e.g. the Page curve). We showed here that this expectation is not correct for the helicity of a massive particle pair. A particle escaping to infinity and its partner particle at the horizon can have all possible combinations of helicity   
\begin{equation}
A_{R\lambda R \lambda_1}\ket{RR}+A_{R\lambda L \lambda_1}\ket{RL}+A_{L\lambda R \lambda_1}\ket{RL}+A_{L\lambda L \lambda_1}\ket{LL}
\end{equation}
where, $R$ and $L$ are the helicities of particles. All the amplitude coefficients can be non-zero.  Entanglement can be even small if the angular distribution of the particle pair is considered. This demonstrates that the helicities of particle pairs are not fully entangled and must be dealt with carefully.  The outgoing Hawking radiation cannot reconstruct all the information of its infalling partners. Since complete entanglement is not true for the vacuum particle pairs, the models that are based on this assumption may not be reliable\cite{Page:1993wv}. 

A more straightforward example is momentum. Momentum is conserved, so two particles may have opposite momenta if they are created by a static object (Fig. \ref{momentum}). Their momenta are correlated, and one particle's momentum is enough to reconstruct its partner's momentum. However, this correlation relation does not preserve after particles propagates a while, because external interactions are involved. The momentum conservation must include the nearby black hole to keep the total momentum conserved.   

This study implies that some of the quantities that are entangled at the beginning may not remain entangled after the particles have propagated away from the initial location. Entanglement is destroyed by interactions.

\begin{acknowledgments}

D.C. Dai is supported by the National Science and Technology Council (under grant no. 111-2112-M-259-016-MY3).  

\end{acknowledgments}

\end{document}